\documentstyle[12pt]{article}

\begin{document}
\def\beq{\begin{equation}}  
\def\eeq{\end{equation}}
\def\beqa{\begin{eqnarray}}
\def\eeqa{\end{eqnarray}}
\def\noin{\noindent}
\def\grad{{\bf \nabla}}
\def\pa{\partial}
\def\pat{{\partial_t}}
\def\rta{\rightarrow}
\def\lra{\leftrightarrow}
\def\cX{{\cal{ X}} }
\def\cA{{\cal{A}}}
\def\cB{{\cal{B}}}
\def\pd{{\partial}}
\def\xia{{\xi}^{\alpha}}
\titlepage
\begin{flushright} QMW-PH-97-19
\end{flushright}
\vspace{3ex}
\begin{center} \bf
{\bf FINITE SOFT TERMS IN STRING COMPACTIFICATIONS WITH BROKEN SUPERSYMMETRY}\\
\rm
\vskip 1.5cm
Imran Shah $\footnote{e-mail: i.shah@qmw.ac.uk} $ and
Steven Thomas $\footnote{e-mail: s.thomas@qmw.ac.uk} $\\
\vspace{2ex}
{\it Department of Physics\\
Queen Mary and Westfield College\\
Mile End Road\\
London E1 4NS\\
U.K.}\\
\vspace{4ex}
ABSTRACT
\end{center}
\noindent
We consider the role of supersymmetry breaking soft terms that are present 
in generalized Narain compactifications of heterotic string theory,
in which local supersymmetry is spontaneously broken, with gravitino
masses being inversely proportional to the radii of compact dimensions.
Such compactifications and  their variants, are thought to be the
natural application of the Scherk-Schwarz mechanism to string theory.
In this paper we show that in the case where this mechanism leads to
spontaneous breaking of $ N=4, d=4 $ local supersymmetry, the limit 
$ \kappa \rta 0 $ yields a 2-parameter class of distinct, dimension $\leq 3$
explicit soft terms whose precise form are shown to preserve the ultraviolet
properties  of $N=4$ Super Yang-Mills theory. This result is in broad
agreement with that of the field theory Scherk-Schwarz mechanism, as
applied to $N=1, d=10 $ supergravity coupled to Super Yang-Mills, although
the detailed structure of the soft terms are different in general. 
      
\newpage
Compactifications of superstring theories to four dimensions, in which 
supersymmetry is spontaneously broken with gravitino masses  inversely proportional
to the radii  of internal tori, have been studied by many authors 
\cite{GDR}, \cite{GDRN}. Such theories
are thought to provide a stringy realization of the well known Scherk-Schwarz
mechanism for spontaneously breaking (local) supersymmetry through a generalized
dimensional reduction (GDR) procedure \cite{S-S}. They are particularly 
interesting in the string context, because there, the number of known supersymmetry 
breaking mechanisms is relatively small. Although historically it has been difficult
to construct realistic low energy models in this way- in particular the 
difficulty concerning the so called decompactification problem associated with the 
need for large radii of compactification, there has been recent progress in this area
\cite{DEC1}, \cite{DEC2}. In any case such a mechanism  can  provide us with a means of breaking supersymmetry when 
stringy mechanisms like compactification on orbifolds are not obviously applicable.
A particular example of this is the recent application of the GDR  mechanism
in compactifying M-theory to four dimensions \cite{M1}. Interestingly in this case, 
a connection has been argued to exist between this mechanism and supersymmetry breaking by gluino condensation  \cite{M2}. 

     In this paper we want explore further the connection between the standard GDR
mechanism as applied to field theories, and its stringy counterpart. We will do this by 
focussing on a rather interesting  property of the mechanism in the context of $N=1, d=10 $
super Yang-Mills (SYM) theory coupled to supergravity compactified to $d=4$
\cite{SOFT1} . In the 
latter paper it was shown how the explicitly broken $N=4$ SYM theory obtained in the 
global limit $\kappa \rightarrow 0 $ where gravity decouples, had the remarkable property
that the supersymmetry breaking terms were precisely of the form that the authors of 
\cite{SOFT2} had shown preserved the  ultraviolet  finite properties of the (unbroken) $N=4$ SYM theory.
These finiteness preserving soft terms were described by a 2-parameter set of masses and 
certain cubic scalar interactions. 

Here we want to see if similar results can be obtained in the string case, in particular 
to the mechanism as applied to Narain type compactifications \cite{NAR}. Such an analysis 
will certainly help further understand the connections between field and string theory 
versions of GDR, because  although they  share a number  of common features, one should 
be cautious in assuming just how far they are in agreement. Indeed our result will show there
is a degree of difference between the two, even though  the general conclusion that 
GDR (in the context of $N=4$ supersymmetry )
produces finiteness preserving soft terms will be verified.  

 Although one could
 presumably carry out such an investigation using a number of different formulations of 
stringy GDR (see refs \cite{GDR} ), we find it attractive to work in the bosonic formalism
constructed in \cite{GDRN} which has close connections with the original Narain construction. 
We begin then, by introducing a number of relevant formulae appropriate to 
a discussion of generalized Narain compactifications of the heterotic string
with spontaneously broken supersymmetry. The reader is refered to \cite{GDRN} for 
further details of the construction. The (light cone) world sheet degrees of freedom are:
non-compact (spacetime) transverse coordinates $X^a , a = 1,2 $;
left/right moving compact coordinates $\bar{X}^i , X^i,  i = 1,.. 6$, sixteen left
moving coordinates $\bar{Y}^I, I = 1,.. 16$ together with four right moving
complex, compactified ``NSR'' bosons $H^A , A = 1, 4 $ which are equivelant
to eight right moving NSR fermions. As, usual the spectrum of the theory 
is obtained from the form of the (right/left moving) Virasosro generators:

\beqa\label{eq:1}
 {L_0} & = & \frac{1}{4} p^2 + {L'}_0 + N - \frac{1}{2} \cr
&& \cr
{\bar{ L}}_0  & = &  \frac{1}{4} p^2 + {{\bar{ L}}_0}' + {\bar N} - 1
\eeqa
where $p$ is non compact spacetime momentum, $ N, \bar{ N}$  number
 operators for right and left moving  oscillators of all types. 
For purely metric compactifications (i.e. the gauge Wilson lines and 
background antisymmetric tensor fields are set to zero which for
simplicity, is the only case we shall consider in this paper)
the compact zero mode contributions ${L'}_0 $ and ${\bar{ L}_0}' $ take the form

\beq\label{eq:2}
 {L'}_0  =  \frac{1}{4} p^t_R  g^{-1} p_R + \frac{1}{2}
 {\tilde p}^t_R h^{-1}
 {\tilde p}_R, \quad \quad 
\bar{ L'}_0  =  \frac{1}{4} p^t_L  g^{-1} p_L + \frac{1}{2}
 {\tilde p}^t_L  c^{-1} {\tilde p}_L 
\eeq
with $p_R = {\hat m} - g {\hat n},\quad  p_L = {\hat m}+ g {\hat n},\quad {\tilde p}_R
= {\hat r} + {\hat t}, \quad {\tilde p}_L = {\hat l} $. Here all hatted
 quantities are integer valued except the shift vector ${\hat t } = 
(- \frac{1}{2}, - \frac{1}{2}, - \frac{1}{2}, 1) $, which is needed in
 order that the compact NSR boson correctly describes bosonic and fermionic
 states. The constant matrix $h$ corresponds to the lattice metric appropriate for 
the NSR bosons and is given in \cite{GDRN}.

The main idea developed in \cite{GDRN} was to show that the above formalism 
(which at present leads to d= 4 theories with extended, unbroken
supersymmetry), can be generalized to include ``Wilson lines''
$x_i^A $ that couple the right moving fields $H^A$ to the coordinates  
 ${ X}^i $ in much the same way as ordinary gauge Wilson lines couple
the coordinates $X^i $ to $Y^I $.  From now on, unless explicitly stated
to the contrary,  by Wilson lines we mean those of $x^A_i  $.

The heterotic world sheet action is then
\newpage
\beqa\label{eq:2.2}
I &=& \frac{1}{2\pi} \int d \tau d \sigma \{ g_{ij} \pd_{\tilde{a}} X^i \pd^{\tilde{a}} X^j
+ h_{AB} \pd_{\tilde{a}} H^A \pd^{\tilde{a}} H^B \cr
&& \cr
&+& c_{IJ} \pd_{\tilde{a}} X^I \pd^{\tilde{a}} X^J
+ \epsilon^{\tilde{a} \tilde{b}} x^A_i \pd_{\tilde{a}} X^i \pd_{\tilde{b}} H^A \}
\eeqa

where $\tilde{a} = 1,2 $ labels the world sheet coordinates, and 
$h_{AB} $ the metric on the lattice corresponding to the compactified
bosons $H^A $, which is given in \cite{GDRN}. $c_{IJ} $ is the 
Cartan matrix of $E_8 \times E_8 $.  
Note that in the light cone gauge, the 4 complex right moving NSR worldsheet fermions 
$\psi^A (z), A = 1, 4$ are given in terms of the chiral bosons $H^A (z)$ by 
$\psi^A (z) = \frac{\scriptstyle 1}{\scriptstyle \sqrt{2}} e^{i H^A (z) }$. Just as in the case of gauge Wilson lines
\cite{NAR}, the presence of the terms in $x^A_i $ in (\ref{eq:2.2}) lead to shifts in the zero mode
terms in the mode expansions of both $X^i $ and  $ H^A$ . This is a result of the fact that
$H^A $ is subject to the  constraint that it be right moving. Standard application of
 dirac bracket quantization leads to the following $x^A_i $ dependence of the 
zero modes $\dot{q}^A $ and $\dot{q}^i $ in the expansions of $H^A $ and $X^i $

\beq\label{eq:2.4}
\dot{q}^A   =  h^{AB} \, (p_B - x_{Bi} n^i \, ) , \qquad
\dot{q}^i  =  g^{ij} \, ( {p'}_j - x^A_i (p_A - \frac{1}{2} x_{Ak} \, n^k )\, )
\eeq
 
whereas $\dot{q}^I = c^{IJ} p_J $ which appears in the expansion of $X^I $,
 is independent of $x^A_i $. The momenta ${p'}_j , 
p_B $ and $p_I $ are all canonical and hence integer valued. The presence of the shift in 
$\dot{q}^A $ is particularly noteworthy, because as mentioned in \cite{GDRN}, it leads to a 
quantization condition on $x^A_i $ .
 This follows directly from demanding that the 
worldsheet supercurrent  has NSR boundary conditions and is related to 
 the preservation of Lorentz invariance in the light cone gauge.
 $\footnote{\rm This was originally pointed out by I. Antoniadis } $
 
Because of this similarity with gauge Wilson lines, one can, as
in the standard case, express the internal zero mode contributions to the
2-d scaling dimension and spin,
 $H = L'_0 + {\bar{ L}'}_0 $,\quad$ S = {L}'_0
 - \bar{L}'_0 $,
 in a way which makes manifest the dependence of the latter
on the various moduli, including now the parameters $  x_i^ A$:

\beq\label{eq:3}
H =  \frac{1}{2} u^t \chi u, \qquad
S  = -\frac{1}{2} u^t \eta u 
\eeq  
with 
\beqa\label{eq:3.1}
\qquad\qquad\eta & = &  \left( \begin{array}{cccc}
    0 & 1 & 0 &0 \\
    1 & 0 & 0 & 0 \\
    0 & 0 & c^{-1} & 0 \\
    0 & 0 & 0 & - h^{-1} \\
  \end{array} \right) \cr
&&\cr 
&&\cr
 \chi &  =  & \left( \begin{array}{cccc}
    g^{-1} & -g^{-1} b' & 0 & g^{-1} {x}^t h^{-1} \\
    -b'^t g^{-1} & (g - b'^t)g^{-1} (g - b') & 0 & (g - b'^t) g ^{-1} {x}^t h^{-1} \\
    0 & 0 & c^{-1} & 0 \\
    h^{-1} {x}  g^{-1} & h^{-1} x g^{-1} (g - b') & 0 &  h^{-1}+ h^{-1} {x}\\
  \end{array} \right)  
\eeqa                                                                     
                                                                             
where $u^t  = ({\hat m}^t,  {\hat n}^t ,  {\hat l}^t , {\hat r}^t +
{\hat t}^t ) $. In (\ref{eq:3.1}), a non vanishing
antisymmetric tensor field $b $ has been included via  $b' = b -\frac{1}{2}
{x}^t h^{-1} {x }$, just to make the formal similarity between
$x_i^A $ and gauge Wilson lines even more apparent, although from now 
on we set $b = 0 $. The Narain metric
$\eta $ is the standard one except for the addition of the metric.

Following \cite{GDRN} it is convenient to go to a canonical basis where  the
metrics $g, c  $ and $h $ are unit matrices at the expense of defining
(moduli) dependent windings and momentum 

\beq\label{eq:4} 
m = e^*_g \, {\hat m} , \qquad  n = e_g \, {\hat n} , \qquad \omega  = e^*_h
\, {\hat \omega} , \qquad l =  e^*_c  \, {\hat l} 
\eeq

where in (\ref{eq:4} ) $ {\hat \omega} = {\hat r } + {\hat t} $ and 
the matrices $e_g , e_c $ and $e_h $ are chosen so that 
$g = e_g^t \, e_g $; $c = e^t_c \, e_c $; and $h = e^t_h \, e_h $. Here $*$ 
represent transpose inverse of a matrix. In this basis (and now setting 
$b = 0 $)  $H$ and $S$ take on similar forms as in (\ref{eq:3} ) with
vectors  ${\tilde u}$  written in terms of the modified windings and momenta
(\ref{eq:4}) and modified matrix $\tilde{ \chi } $ given by 

\beq\label{eq:5}
\tilde {\chi}   =  \left( \begin{array}{cccc}
    1 & \frac{1}{2} {\cX}^2  & 0 & \cX^t \\
    \frac{1}{2} \cX^2 & (1 + \frac{1}{2} \cX^2 ) &0 &  ( 1 + \frac{1}{2}
    \cX^2 )\cX^t \\
      0& 0 & 1 & 0 \\
    \cX & \cX (1 + \frac{1}{2} \cX^2 ) &0  & 1 + \cX^2 \\
      \end{array} \right)
\eeq
with $  \cX^A_i =  {(e^*_h x e^{-1}_g )}^A_i, \quad  \cX^2 \equiv \cX^t \cX $.

In this basis the mass squared and level matching of physical states can be
shown to be equivalent to 

\beqa\label{eq:6}
\frac{1}{2} M^2  & = & \frac{1}{2} ( m + \cX^t \omega + (1 + \frac{1}{2}
\cX^2 )n )^2 + l^2 + 2 \bar{ N} - 2 \cr
&&\cr
0 & = & - m n - \frac{1}{2} (l^2 - \omega^2 ) + N - \bar{N} + \frac{1}{2}
\eeqa

With $\cX_i^A = 0 $, the spectrum (\ref{eq:6}) describes N=4, d=4
supergravity coupled to super Yang-Mills. Massless spacetime fields
within a supermultiplet correspond to states  with vectors $\omega $ satisfying  $\omega^2 = 1
$. There are eight physical states 
with $ (\omega^t =  \{  (\pm  1,0,0,0) +$
permutations $\} $ describing  bosons, whilst  $\omega^t =  \{ (\pm
\frac{1}{2}, \pm \frac{1}{2}, \pm \frac{1}{2},\pm  \frac{1}{2}) $,  with even
number of minus signs $\} $, describe spacetime fermions. We shall refer to
these sets of lattice vectors as $\omega_B$ and $\omega_F $ respectively. 
 The spectrum (\ref{eq:6}) generally leads to a
spontaneously broken spectrum when $\cX_i^A \neq 0 $.
Indeed it is easy to see that the four  d = 4 gravitini $\psi^\mu_A ,\quad A = 1,
4 $ obtain a mass squared $ ({M^A}_{3/2})^2 $  given by 

\beq\label{eq:7} 
({M^A}_{3/2})^2 =   \omega^{(A)t} \cX   \cX^t \omega^{(A)} ,
\qquad  A = 1, 4, \qquad\{\pm  \omega^{(A)t} \} = \omega_F 
\eeq

where in light cone gauge, the gravitinos correspond to the states

\beq\label{eq:8}
\psi^A_a  \sim \bar{ X}^a_{-1}  \, {e}^{{\scriptstyle i
\omega}^{(A)} \cdot {\scriptstyle H} } | k >
\eeq
$k$ being non-compact, transverse momentum, and it is assumed that 
the vertex operators in (\ref{eq:8}), as elsewhere in this paper are normal
ordered.

At this point one can see that additional constraints have to be placed on 
$\cX$ in order to avoid giving a mass to the graviton, which should
presumabely be related to 2-loop modular invariance. Making the basis
choice that the lattice vectors ${\omega^{(a) t}} = (1,0,0,0), (-1,0,0,0),
a = 2,3 $ correspond to the 2 transverse directions in $d =4 $,
states corresponding to gravitons are of the form
$\bar{ X}^a_{1} \, e^{i \omega^{(b)}\cdot H } | k >  $. It is easy
to show that the condition of vanishing graviton mass then implies that the
first column of $  \cX $ must vanish, so we take 
\beq\label{eq:9}
\cX^{t A}_i =  \left( \begin{array}{cccccc}
    0& 0 & 0 & 0 & 0 & 0  \\
    b_1& b_2 &b_3 &b_4 &b_5 &b_6 \\
    c_1& c_2 &c_3 &c_4&c_5&c_6  \\
    d_1& d_2 &d_3 &d_4&d_5&d_6 \\
  \end{array} \right)  
\eeq
where we remind the reader that in the canonical basis, 
the Wilson line parameters $b_1, ...
d_6 $ are continuous parameters that are functions of the moduli in
general. 

In this paper we want to concentrate on the soft terms appearing in the
 matter sector-i.e. to the adjoint scalars and fermions of N=4 SYM. There
 are 4 majorana fermions and 6 scalars,  which in the present formalism
 correspond to the states

\beqa\label{eq:11}
 &&e^{\scriptstyle  i l \cdot \bar{ Y} } \, e^{ \scriptstyle i \omega \cdot H} | k > , \qquad
\bar{ Y}^I_{-1} \, e^{\scriptstyle  i \omega \cdot H} | k >, \quad I = 1,.. 16 ;
\qquad l^2 = 2, \quad 
 \omega \in \Omega_F    \cr
&&\cr
&& e^{\scriptstyle  i l \cdot \bar{ Y} } \, e^{\scriptstyle i \omega \cdot H} | k > , \qquad
\bar{ Y}^I_{-1} \, e^{\scriptstyle  i \omega \cdot H} | k >, \quad I = 1,.. 16 ; 
\quad l^2 = 2, \quad \omega \in {\Omega}'_B    
\eeqa
where in (\ref{eq:11}), ${\Omega}'_B $ corresponds to the set of 6 
orthogonal vectors $\Omega_B $ defined earlier, but with the vectors 
$ \pm (1,0,0,0) $ excluded. It is useful to divide 
${\Omega}'_B $ into two parts so that 
$ {\Omega}'_B = \{ \omega^{( \alpha )},- \omega^{( \alpha )} \}, \alpha =1,
..3   $ in the same way that  $\Omega_F = \{ \omega^{(A)}, - \omega^{(A)} \}, A =
1,.. 4 $. 

\newpage
The $6 \times 6 $ and $4 \times 4 $ mass matrix squared , $M^2_0 $ and $ M^2_{1/2}
 $ of these states are readily obtained from  (\ref{eq:6}):
\beqa\label{eq:12}
M^2_0 \quad & = & \quad {\rm Diag} \{ \omega^{(\alpha ) t} \cdot \cX {\cX}^t
\cdot \omega^{(\alpha )},\quad  
\omega^{(\alpha ) t} \cdot \cX {\cX}^t \cdot 
 \omega^{(\alpha ) } \} \qquad \alpha = 1,..3 \quad;
\cr
&&\cr
M^2_{1/2} \quad & = & \quad {\rm Diag} \{ {\omega}^{(A )t} \cdot \cX {\cX}^t 
\omega^{(A )}, \quad
 \omega^{( A) t} \cdot \cX {\cX}^t \omega^{(A)}  \} \qquad A = 1,..4 
\eeqa

The two fold degeneracy of scalar masses in (\ref{eq:12}) is indicative
that these masses are of `$ {( A^{\alpha})}^2 + {(B^{\alpha} )}^2 $  ' type,
where $A^{\alpha} $ and $B^{\alpha}$ are  scalar and psuedoscalar fields
respectively. Such scalars  appear as the internal components of $d =10 $
gauge fields.
 
An important test of whether explicit
supersymmetry breaking masses can preserve certain ultraviolet properties
of the unbroken theory, is the vaishing of the graded trace of $M^2 $,
denoted by  ${\rm Gr Tr} M^2 $, 
 where we adopt the conventions that
\beq\label{eq:12.5}
 {\rm Gr Tr}M^2 = 
 \sum_s {(-1 )}^{2 s +1} \,  (2 s +1) \, M^2_{s} 
\eeq 
with $M^2_s $ being the mass squared matrix of the field with spin $s$.

Since we have set all the  gauge Wilson lines to zero, there is no gauge
symmetry breaking and so only the adjoint scalars and spinors contribute 
to ${\rm Gr Tr} M^2 $ over the Yang-Mills multiplet.
One can now easily check by explicitly substituting the matrix  $\cX $ into the 
mass formula (\ref{eq:12}) that indeed ${\rm Gr Tr} M^2 = 0 $ identically.
It is interesting to note, by for example relaxing the graviton constraint 
(\ref {eq:9}), that the graded trace only vanishes 
when this condition is implemented. In passing it should be mentioned 
that the vanishing of ${\rm Gr Tr} M^2 $ and its generalizations 
has been shown to occur within the framework of 
`misaligned supersymmetry' in string theory,
 considered in \cite{misaligned}. It would be interesting to see 
explicitly how the stringy GDR mechanism fits into this. 

 At this point it is useful to  remind the reader some of the results concerning
the general structure of soft terms in $N=4$ SYM that preserve finiteness
\cite{SOFT2}.
Some years ago superspace spurion techniques were employed to address this
problem, and the conclusions were as follows.
Let $A_{\alpha}, B_{\alpha} , \Psi_{\alpha},\lambda,\quad \alpha = 1,.. 3 $ and
$A_{\mu} $ be the field content of N=4 SYM. This choice of field 
decomposition is natural when writing the theory in terms of
$N=1$ superfields, where the scalars/psuedoscalars $A_{\alpha}, B_{\alpha} $
and Majorana fermions $\Psi_{\alpha} $ lie in $N=1$ chiral multiplets and 
the gaugino $\lambda $ and gauge fields $A_{\mu } $  in $N=1$ vector
multiplets. All fields transform in the adjoint representation of 
the gauge group $G$. 

The authors of \cite{SOFT2}  considered the following explicit supersymmetry breaking
mass terms:

\beqa\label{eq:14}
S^{(1)}_{mass} & \equiv & \frac{\scriptstyle 1}{ \scriptstyle C(G)}\int d^4 x \,{\rm Tr}\,  \{ \frac{1}{2} U_{\alpha 
\beta }\, (A_{\alpha}A_{\beta} +  B_{\alpha} B_{\beta} ) \} \cr 
&& \cr
S^{(2)}_{mass} & \equiv & \frac{\scriptstyle 1}{\scriptstyle C(G)}\int d^4 x \,{ \rm Tr}\,\{ \frac{1}{2} V_{\alpha \beta
}\, ( A_{\alpha} A_{\beta} -  B_{\alpha} B_{\beta} ) \} \cr
 && \cr
S^{(3)}_{mass} & \equiv & \frac{\scriptstyle  1}{\scriptstyle C(G)} \int d^4 x\,{\rm Tr} \,\{  M_{\alpha \beta}\,
\bar{\Psi}_{\alpha}{\Psi}_{\beta} + {\cal{M}  }
 \bar{\lambda  } \lambda \} \ 
\eeqa

In (\ref{eq:14}),  $V_{\alpha \beta} $ are `$ A^2 - B^2 $'  masses that 
are known to preserve finiteness \cite{SOFT2} . Such masses of course do not
contribute to  the  ${\rm Gr Tr} M^2 $. The remaining mass parameters
$U_{\alpha \beta }, M_{\alpha \beta} $ and $\cal{M} $ are subject to the
single condition that ${\rm Gr Tr} M^2 = 0 $. This condition is a
necessary, but not sufficient condition to preserve the  finiteness of the
expicitly broken $N=4 $ theory. In \cite{SOFT2} it was shown that divergences
cancel if in addition to the general mass terms  discussed above, fixed 
cubic scalar interaction terms are added, with couplings related to the 
fermion masses above. That is,  one must add

\beqa\label{eq:15}
S^{(1)}_{cubic} & = & \frac{\sqrt{2}}{\scriptstyle  C(G)}\int d^4 x \,{\rm Tr} \, \{ 
N_{\alpha \beta \gamma} \, ( A_{\alpha} ( A_{\beta} A_{\gamma} -  B_{\beta}
 B_{\gamma} )  + 2 B_{\alpha } A_{\beta} B_{\gamma} )  \} \cr 
&&\cr
S^{(2)}_{cubic} & = & \frac{\sqrt{2}} {3! \scriptstyle  C(G)} \int d^4 x \,{\rm Tr}\, \{ 
 N_0  \, {\epsilon}_{\alpha \beta \gamma} ( A_{\alpha}  A_{\beta} A_{\gamma} 
  -3  A_{\alpha } B_{\beta} B_{\gamma} )  \} 
\eeqa 
 
where the parameters $N_{\alpha \beta \gamma }$ and $N_0 $ are given 
in terms of the fermion masses by

\beq\label{eq:16}
N_{\alpha \beta \gamma } =  \frac{  i g}{\scriptstyle  2 \sqrt{2}} M_{\alpha \delta }
\,{\varepsilon}_{\delta \alpha \beta} ; \qquad
N_0 =  \frac{ i g}{\scriptstyle  2 \sqrt{2}} \cal{M} 
\eeq
and in (\ref{eq:15}) and (\ref{eq:16}), ${\rm Tr} $ 
indicates the trace
over the adjoint representation of the gauge group $G$, with $ C(G)$
the quadratic Casimir, and $g$ the gauge coupling. We choose a
normalization  of the group generators $t^x$ so that
 ${\rm Tr} ( t^x t^y t^z ) = (i/2) C(G) {f}^{xyz}, \quad
x= 1, ... dim G \, $, and the coupling  $g$ is chosen with the standard
normalizations of Yang-Mills terms, (the coupling constant in \cite{SOFT2} is 
$\sqrt{2} $ times ours ).

The fact that the soft masses arising from the generalized Narain
compactifications satisfy graded sum rules is further
evidence that this constuction is the natural application of the
Scherk-Schwarz mechanism to superstring theory. 
Having at least shown that one of the conditions for finiteness preserving
soft terms discussed above, holds, we now want to consider 
the role of cubic scalar interactions.

The problem we want to address now is whether such scalar cubics can appear
at all in the generalized Narain compactification we are considering here,
and secondly to what extent are they the ones required by finiteness of the
explicitly broken low energy N=4 SYM formally obtained when we take the 
limit $\kappa \rightarrow 0 $. To this end we need to calculate a tree level
string scattering amplitude involving three of the $(d =4 )$ adjoint scalar
fields $A_{\alpha} , B_{\beta} $. Since these fields are obtained as the 6 internal
components of the 10-dimensional gauge fields $A_{\tilde{\mu}}, \tilde{\mu} = 0, ..9 $,
it is natural to look at the (massless) vector boson emission vertex as a starting
point in constructing the (adjoint) scalar emission vertex in the heterotic string.
Recall that the vertex describing the emission of an on-shell state with polarization 
$\Xi^{\tilde{\mu}}(K)$
and momentum $K^{\tilde{\mu}}$, \, ( $ K^2 = 0$, and $K\cdot \Xi = 0 $ ), $V_B (\Xi, K ) $
is given in the NSR formalism as \cite{GSW}

\beq\label{eq:17}
V_B = \frac{  g}{\scriptstyle  3 C(G)}
\{ \, \Xi \cdot {\dot{X}} - \Xi \cdot \psi \, K \cdot \psi \,  \} \,  e^{i K \cdot X}
\eeq
where in (\ref{eq:17}) we have supressed the left moving vertex operators that describe 
the gauge quantum numbers  of the vector boson. Inclusion of these, and the mass shell 
condition $K \cdot K = 0 $ enures that $V_B $ is a physical $(1,1) $ operator.
 The normalization
factor in $V_B$ is needed to reproduce the standard Yang Mills kinetic term 
$-1/(4 g^2 {\scriptstyle  C(G)} )\int d^{10} x \, {\rm Tr}\, ( F^2 ) $.
The 6 internal polarizations 
$\xi^i \equiv \Xi^i ,\,  i = 4, .. 9 $, describe adjoint $d=4$ scalars,
if we include the effects  of the left moving $E_8 \times E_8 $ lattice degrees
of freedom by taking $\xi^i $ to transform in the adjoint representation 
of this group. Decomposing the momentum 
$K^{\tilde{\mu}} = \{ k^\mu, p^i \} $ it is easy to see that the 3-scalar scattering
amplitude appropriate to standard Narain compactifications is
\beq\label{eq:18}
<k_1, p_1, \xi_1| V_B(k_2,p_2,\xi_2 )|k_3,p_3, \xi_3> =
\frac{  g}{\scriptstyle  3 C(G)} {\rm Tr} \{ \xi_1 \cdot \xi_3 p_3 \cdot \xi_2 +   
 \xi_2 \cdot \xi_1  p_1 \cdot \xi_3 + \xi_3 \cdot \xi_2 p_2 \cdot \xi_1 \}
\eeq
Now it is clear from this that the mass shell condition  $ k^2 = 0,\, p^2 = 0 $
appropriate for toroidal compactification, implies the amplitude (\ref{eq:18}) vanishes,
which is consistant with the fact that unbroken $N=4, d=4$ supersymmetry  forbids the existence
of such cubics. It follows from this that mechanisms that break any of these supersymmeries 
could generate non vanishing scalar cubic interactions.
Thus to compute  adjoint scalar emission vertex when $x^A_i \neq 0 $ we 
should take a lead from the gauge boson emission vertex $V_B $ given in
 (\ref{eq:17}). In particular the emission vertex $V_B $ has the property
    that \cite{GSW}
\beq\label{eq:19}
\{ G_r , V_0 (\Xi , K ) \} = V_B (\Xi , K ), \quad V_0 (\Xi , K ) 
= \Xi \cdot   \Psi \,  e^{i {K} \cdot { X}} , \quad r = \pm \frac{1}{2}, 
\pm \frac{3}{2} , ....
\eeq
where $G_r $ are the oscillators occuring in the expansion of 
right-moving  world sheet supercurrent $J_z (z) $
in the  NS sector (once again we supress the left moving contributions in
 (\ref{eq:19}) ).
In the present formalism generator  $J_z (z) $ is given, in the light cone gauge, by
\beq\label{eq:20}
J_z (z) =  \pd_z \bar{X}  {e}^{\scriptstyle  i H^1} + \pd_z
X \, {e}^{\scriptstyle  -i H^1} + \sum_{\alpha=1}^3 
( \pd_z X^{\alpha} \, {e}^{\scriptstyle  -i H^{\alpha +1}} + \pd_z
\bar{X}^{\alpha}  \,   {e}^{\scriptstyle  i H^{\alpha +1}} )  
\eeq
with the definitions that $X = \frac{1}{\sqrt{2}} ( X^{\mu = 2 } + 
i X^{\mu = 3} ) ; $\,$ X^{\alpha} = \frac{1}{\sqrt{2}} ( X^{i= 2 \alpha+2} + 
i X^{i = 2 \alpha + 3} )$

The property of the gauge boson emission vertex (\ref{eq:19}) is
essentially a
consequence of unitarity. The vertex $V_0 (\Xi, K ) $ creates 
conformal dimension $\frac{1}{2} $ states when acting on the vacuum, when
one imposes the mass shell condition.
The emission vertex of the $( d=4 ) $ adjoint  scalars that originate
from such $d = 10 $ gauge fields should be given by a similar construction,
except we have to include the effects of having $x^A_i $ non vanishing.
In particular, the analogue of $V_0 (\Xi , K ) $ for these scalars,
which we denote $\tilde{V}_0 (\xi, \bar{\xi}, k,\bar{k}, x ) $ is 

\beq\label{eq:21}
\tilde{V}_0 (\xi, \bar{\xi}, k, \bar{k}, x ) = ( \xia \,  {e}^{\scriptstyle
- i \omega^{ (\alpha )}
\cdot H } + {\bar{\xi}}^{\alpha} \, {e}^{\scriptstyle i \omega^{ (\alpha )} \cdot H } )
\,\, {e}^{\scriptstyle i (\, {\bar{k}}^{( \alpha )} X + k^{(\alpha )}
\bar{X} \, )}
\eeq
where we have defined complex polarizations (corresponding to wavefunctions
of the adjoint scalars)
 $\xia =  \frac{\scriptstyle  1}{\scriptstyle  \sqrt{2}} ( \xi^{i= 2 \alpha+2} + 
i \, \xi^{i = 2 \alpha + 3} )$.   

One can easily check that the mass shell condition for the 6 adjoint scalars 
namely ${\bar{k}}^{( \alpha )} k^{( \alpha ) } = - {( \cX^t \cdot \omega^{( \alpha
)} )}^2  $ (c.f.(\ref{eq:12})), together with the shifted  momentum
$\dot{q}^i $ of (\ref{eq:2.4}) implies that states created with 
$\tilde{V}_0 $ have (right- moving) conformal dimension $h = \frac{1}{2} $.
(Including the $E_8 \times E_8 $ lattice degrees of freedom in $\tilde{V}_0
$ would give $\bar{h} = 1 $ ). The correct  dimension $(1,1) $ 
emission vertex  $V_S (\xi , \bar{\xi}, k, \bar{k}, x ) $ is then given 
by $\{ G_r , \tilde{V}_0 \} $, where $G_r $ are the modes 
 appearing in the right moving supercurrent with $x^A_i $ non vanishing.
In calculating $V_S $ one has to take care about the shifts in the 
zero modes ${\dot{q}}^{\alpha}, {\bar{\dot{q}}}^{\alpha} $ appearing in 
$X^{\alpha} $ and ${\bar{X}}^{\alpha} $. After some algebra one finds the
following expression for the (on-shell) vertex $V_S $  

\beqa\label{eq:22}
V_S (\xi ,\bar{\xi}, k, \bar{k}, \cX , \bar{\cX} ) & = & \frac{  g}{\scriptstyle  3 C(G)}
\sum_{\alpha = 1 }^3 \{ \, \xia  {\bar{\dot{ X}}}^{\alpha} + 
\bar{\xia}  { \dot{ X} }^{\alpha} +  \sum_{\beta}\,  [ {\bar{\cX}}^{t \beta}
\cdot \omega^{(\alpha )} ( \psi^{\beta} \bar{\psi}^{\alpha } \xia \cr
&&\cr
-  \psi^{\beta} \psi^{\alpha} {\bar{\xi}}^{\alpha} )& -& 
 {\cX}^{t \beta} \cdot \omega^{(\alpha )} ( 
{\bar{\psi}}^{\beta} {\bar{\xi}}^{\alpha}  - {\bar{\psi}}^{\beta}
{\bar{\psi}}^{\alpha} \xi^{\alpha}\,  ) \, ]\,  \} \, e^{i {\bar{k}}^{(\alpha )} X
 + i k^{(\alpha )} {\bar{X}} } \cr
&& \cr
 {\rm with} &\psi^{\alpha} & = \frac{1}{\displaystyle \sqrt{2}} \, {e}^{\scriptstyle  i \omega^{(\alpha )} \cdot H }
\eeqa

In (\ref{eq:22}) we have defined complex Wilson lines
 $\cX^A_{\alpha} = \frac{1}{\sqrt{2}} ( \cX^A_{i = 2 \alpha +2} +  i 
 \cX^A_{i = 2 \alpha + 3} ) $ and we remind the reader that the lattice
 vectors $\omega^{( \alpha )} = (0,1,0,0) $ etc.
As it stands the emission vertex $V_S$ as defined in (\ref{eq:22})
has a normal  ordering problem concerning the bilinear terms in the 
vertex operators  $ e^{i \omega^{(\alpha )} \cdot H } $ (or equivalently,
 the fermions $\psi^{\alpha} $). On the face of it there is a similar
 ordering problem concerning the gauge boson vertex $V_B$ (\ref{eq:17}).
However there, as is well known, the mass shell condition $K\cdot \Xi = 0 $
projects out possible pole terms and there is no ordering problem.
In the case of $V_S $ however, the mass shell condition by itself is not
 sufficient and one has to impose further conditions on the Wilson lines.
These conditions can be readily obtained

\beq\label{eq:23}
\cX^{t \alpha } \cdot \omega^{ ( \alpha ) } = 0 ; \quad 
{\bar {\cX}}^{t \alpha} \cdot \omega^{ (\alpha ) } = 0 , 
\quad \alpha = 1..3 
\eeq
i.e the diagonal elements of the ( $3 \times 3 $  ) complex matrix
$ \cX^{t \alpha }\cdot \omega^{ (\beta ) } $ are required to vanish.
 
This constraint, together with the one imposed by massless gravitons 
(\ref{eq:9}) leads to the following form for the (real) Wilson lines
$\cX^A_i $

\beq\label{eq:24}
 { \cX^{t A}_i}   =   \left( \begin{array}{cccccc}
    0& 0 &0 &0 &0& 0 \\
    0& 0 &b_3 &b_4&b_5&b_6 \\
    c_1& c_2 &0 &0&c_5&c_6\\
    d_1& d_2 &d_3&d_4&0&0 \\
  \end{array} \right)  
\eeq 
    
Armed  with  $V_S $ it is staightforward to calculate the amplitude 
\beqa\label{eq:24.5}
A_{cubic} & = & 
<k_1, \bar{k}_1, \xi_1, \bar{\xi}_1 | V_S ( k_2, \bar{k}_2 , \xi_2 ,
\bar{\xi}_2 , \cX , \bar{\cX} ) | \xi_3 , \bar{\xi}_3 , k_3 , \bar{k}_3 > \cr 
&& \cr
| \xi , \bar{\xi} , k , \bar{k} >  &= & \xia | {\omega}^{(\alpha )} , k
 , \bar{k} > + {\bar{\xi}}^{\alpha} | - {\omega}^{(\alpha )} , k
 , \bar{k} >
\eeqa 
Including the $E_8 \times E_8 $ degrees of freedom, 
one obtains  the following explicit supersymmetry breaking 
trilinear soft terms in the effective action 

\beq\label{eq:25}
 S_{cubic} = \frac{ -4 g i}{\scriptstyle 3 C(G)} \int d^4 x {\rm Tr} \,\sum_{\alpha ,\beta =
1}^{3} \{ \frac{3}{2} {\cal{M}}_{R}^{\beta \alpha}   A^{\alpha} B^{\beta} 
B^{\alpha} + \frac{1}{2} {\cal{M}}_{I}^{\beta \alpha}   A^{\alpha} A^{\beta} 
B^{\alpha} \}
\eeq

where the adjoint (psuedo) scalars ( $B^{\alpha} $ ) $ A^{\alpha } $  are
defined in terms of the polarizations $\xia $ by   $\xia = B^{\alpha} + i A
^{\alpha} $. Note that the net result of including the gauge degrees
of freedom result is to add an adjoint group index to the polarizations. The mass matrices 
${\cal{M}}_{R}^{\beta \alpha} $ and ${\cal{M}}_{I}^{\beta \alpha} $ are
given by 

\beq\label{eq:26}
 {\cal{M}}_{R}^{\beta \alpha}  =  Re ( {\cX}^{t \beta} \cdot 
\omega^{(\alpha )} ) \quad, \quad 
{\cal{M}}_{I}^{\beta \alpha}  =  Im ( {\cX}^{t \beta} \cdot 
\omega^{(\alpha )} )
\eeq

From the structure of these soft terms, we can see that the 
$  {\cal{M}}_{R}^{\beta \alpha} $ and $  {\cal{M}}_{I}^{\beta \alpha}$
are parity even (odd) repectively. Having obtained an explicit realization of
the soft cubics appearing in generalized Narain compactifications, 
the final step is to  check whether these really are the ones necessary to 
preserve finiteness of broken $N=4 $ SYM. From our earlier discussion, 
to do this we need to have an explicit form of the   fermion mass matrix.
Note that although we know the form of the mass squared matrix (c.f.(\ref{eq:12})) the mass matrix requires a bit more work to 
get in a form that depends linearly on $\cX $. In fact we shall see that there are
two (complementary) approaches to deriving this matrix. 
The first approach is simply to consider the effective $d = 4 $ dirac
equation for the fermions in the $N= 4 $ SYM multiplet, when supersymmetry
is broken by $\cX^A_i $. As we have seen (\ref{eq:2.4}), the latter
parameters generate shifts in the 6 dimensional internal momentum, and one
can verify that in this event, the induced  $4 \times 4 $
fermion mass matrix is 

\beqa\label{eq:27}
 {M_{1/2}}_{AB} & = &  \sum_{\delta = 1}^3 {\cal{B}}^{\delta}_{AB}
 \,\, {\rm Re}\, ( \cX^{t \delta } ) \cdot \omega^{(B)} - 
  (\,  \, \sum_{\delta = 1}^3 {\cal{A}}^{\delta}_{AB}
 \,\, {\rm Im}( \cX^{t \delta} ) \cdot \omega^{(B))} \,)  \, \gamma_5 \cr
&&\cr
& \equiv & {M_{1/2}^{(+)}}_{AB} + {M_{1/2}^{(-)} }_{AB} \, \gamma_5
\eeqa
where the 6 matrices ${\cal{A}}^{\delta}_{AB} $ and ${\cal{B}}^{\delta}_{AB} $
are $SU(2) \times SU(2) $ generators, an explicit representation of which
is given in \cite{GSO}. In obtaining (\ref{eq:27}) we have used the 
representation of the $ d= 10 $ dirac algebra given in \cite{GSO} and the 
properties

%


\beq\label{eq:29}
\{ \cA^{\alpha} , \cA^{\beta} \} = 
\{ \cB^{\alpha} , \cB^{\beta} \} = -2 {\delta}^{\alpha \beta}, \quad 
[ \cA^{\alpha},  \cB^{\beta} ] = 0
\eeq


 
Again we see that it is the real and imaginary parts of the complexified
Wilson line $\cX^A_{\alpha} $ that generates parity even and odd 
terms -this time via fermion masses. 

Now although one might expect that squaring the fermion mass matrix 
in (\ref{eq:27}) would give the mass squared
formula (\ref{eq:12}) in fact one does not in general reproduce this. 
Defining the  6 diagonal matrices,  ${\Sigma_{\alpha}}_{AB} = 
 {\rm Re}\, ( \cX^{t}_ {\delta } ) \cdot \omega^{(B)} \, {\delta}_{AB} 
 $ and $ {{\Sigma'}_{\beta}}_{AB} = - {\rm Im}\, ( \cX^{t}_ {\alpha } )
 \cdot \omega^{(B)} \, {\delta}_{AB} $, i.e. 
\beqa\label{eq:31}
{\Sigma_1}_{AB}  &=& {\rm Diag}( -c_2-d_2, c_2-d_2,-c_2+d_2, c_2 +d_2) \cr 
{\Sigma_2}_{AB} &=& {\rm Diag}( b_4-d_4, -b_4-d_4,-b_4+d_4, b_4 +d_4)\cr
{\Sigma_3}_{AB}  &=& {\rm Diag}( b_6-c_6, -b_6+c_6,-b_6-c_6, b_6+c_6 )\cr
{{\Sigma}'_1}_{AB}  &=& {\rm Diag}( -c_1-d_1, c_1-d_1,-c_1+d_1, c_1 +d_1)\cr
{{\Sigma}'_2}_{AB} &= &{\rm Diag}( b_3-d_3, -b_3-d_3,-b_3+d_3, b_3 +d_3)\cr
{{\Sigma}'_3}_{AB}  &=& {\rm Diag}( b_5-c_5, -b_5+c_5,-b_5-c_5, b_5+c_5 )
\eeqa

then one finds

\beq\label{eq:32}
M_{1/2}^2 = {( \cB \cdot \Sigma )}^2 + {( \cA \cdot {\Sigma '} )}^2
          +i \gamma_5 \, ( ( \cB \cdot \Sigma )(\cA \cdot {\Sigma}' )
- (\cA \cdot {\Sigma}') (\cB \cdot \Sigma ) ) 
\eeq

whereas (\ref{eq:12}) implies, (after some algebra) that  $ M_{1/2}^2 = \Sigma \cdot \Sigma + 
\Sigma ' \cdot \Sigma ' $
From these definitions, it can be seen that $\Sigma_{\alpha} $ and
 ${{\Sigma}'}_{\alpha} $
induce parity even and odd soft terms  respectively.

There is another way of viewing the above ``mismatch'' between 
(\ref{eq:31}) and (\ref{eq:12}). It follows from considering the
superconformal algebra in the R sector where one expects that 
$\{ G_0, G_0 \} = L_0 $. If one focuses on the internal contributions to 
$G_0 $ and in particular noting the various shifts in momenta induced by
$\cX^A_i $, then one again finds an obstruction to the anticommutator of 
$G_0 $ giving  the correct eigenvalues of $L_0 $, when say we act on the
sates corresponding to the adjoint  fermions considered above. 
Thus we have to impose further conditions on the Wilson line parameters
in order that  either the superconformal algebra is satisfied or
equivelantly  that $M_{1/2}^2 $ agrees with (\ref{eq:12}). It is 
sufficient to impose 

\beq\label{eq:33}
\{ \cB^{\alpha}, \Sigma_{\beta} \} = 0 \quad, 
\{ \cA^{\alpha}, {\Sigma}'_{\beta} \} = 0 \quad,
[ \cB\cdot \Sigma , \, \cA \cdot \Sigma ' ] = 0
\eeq
where it is understood that the indices $\alpha , \beta $ in 
(\ref{eq:33}) run over only those values for which the corresponding 
$\Sigma_{\alpha} $ and ${{\Sigma}'}_{\alpha} $ are non vanishing.

Using the explicit representation of the matrices $\cA^{\alpha} $ 
and $\cB^{\alpha} $ given in \cite{GSO}, we have found a class of solutions where 
only one of each matrix $\Sigma_{\alpha} $ and ${{\Sigma}'}_{\alpha} $ 
as listed in (\ref{eq:31}) is taken to be non
vanishing at a time. This corresponds then to solutions where only 2 of the
12 parameters are  turned on at a time. However such solutions are not
strictly independent. For example if we take the solution where $\Sigma_3 \neq 0 $,
 all others vanishing, then one can show that the two other solutions where 
$\Sigma_1 $ and $\Sigma_2 $ are turned on separately produce isomorphic theories, in
that the fermion mass eigenvalues, scalar masses and trilinear scalar
couplings are all equal up to a permutation of the 
$\alpha $  indices on the fields $A_{\alpha} , \, B_{\alpha} $ 
 respectively. A similar argument
applies to the parity odd solutions involving non vanishing 
 ${\Sigma '}_{\alpha} $, in which there is again only a 2-parameter set of distinct
solutions. Therefore the result is that we have a (mutually exclusive) 
2-parameter set of parity even and 2-parameter set of parity odd soft
terms respectively.

Now we are in a position to finally check if the cubic soft terms 
appearing in (\ref{eq:25}) are indeed of the precise form needed for 
finiteness. Note that the original results by the authors of \cite{SOFT2} 
did not include parity odd soft terms, which we have seen are generated in 
the generalized Narain compactifications considered here. 
Thus for the present we set $ {{\Sigma}'}_{\alpha} = 0 $ and consider parity even terms
only.  

Consider then the solution to (\ref{eq:33}) where we take $\Sigma_3 
\neq 0 $ only. The fermion  mass matrix is then

\beq\label{eq:34}
M_{1/2} = \frac{1}{2}  \left( \begin{array}{cccc}
    0&b_6-c_6&0&0\\
    b_6-c_6&0&0&0 \\
    0&0&0&-(b_6 +c_6)\\
    0&0&-(b_6+c_6)&0\\
  \end{array} \right)
\eeq
Diagonalizing gives $M_{1/2}^{diag} = \frac{\scriptstyle 1}{
\scriptstyle 2}(b_6-c_6, -b_6+c_6 ,b_6+c_6,-b_6-c_6)$
where we identify the gaugino mass with $- \frac{\scriptstyle 1}{
\scriptstyle 2 } (b_6 +c_6 )$. Note that in this
diagonalization procedure, the labels $\alpha =1..3$ on the scalar and
psuedoscalar fields are rotated compared to those before diagonalization.
This is easily seen from checking how the yukawa couplings of  $A_{\alpha}$
and $B_{\alpha}$ change under the orthogonal transformation that
diagonalizes $M_{1/2} $. Taking this into consideration,  one finds the
cubic interactions (\ref{eq:25}) expressed in terms of 
 the (rotated) fields is 

\beq\label{eq:35}
S_{cubic} = \frac{4 g i }{\scriptstyle  C(G)} \int d^4 x \, {\rm Tr} \, \{ c_6 A_2 B_3 B_1 - 
b_6 A_1 B_3 B_2 \}
\eeq    

It can now be checked, using the diagonalized mass matrix $M_{1/2} $ that
these agree precisely with those required by the finiteness 
conditions (\ref{eq:15}) and (\ref{eq:16}). As we mentioned, there are 
also parity violating solutions to (\ref{eq:33}) which correspond to 
parity violating soft terms. These have very similar structure to the 
parity preserving ones, but their effects on the finiteness of 
$N=4 $ SYM does not seem to have been studied in the literature.
It is tempting to speculate that they will turn out to be 
finiteness preserving as well.  
 
We can make a number of observations about these results. First it is
interesting to compare the soft terms we have obtained here in the
context of string compactification, and those obtained by applying the 
standard field theory mechanism to $N= 1, d=10 $ SYM coupled to
supergravity \cite{SOFT1}. In the latter, a 2 parameter set of
(parity even) cubic scalar
soft terms and soft masses were also obtained, that preserved the
finiteness constraints. Although the form of these terms shows they are 
roughly similar -they are not identical. In fact there are good reasons why
they cannot be in general. The structure of the cubics found
in (\ref{eq:35}) shows that  terms trilinear in $A_{\alpha} $ 
are not present. From a calculational point of view, this absence
was due to the fact that no third rank antisymmetric tensor exists 
in the formalism, built from $\cX^A_{\alpha} $, which is needed in order 
to couple to a term like ${\rm Tr}\, \{  A_{\alpha}A_{\beta} A_{\gamma} \} $ which
is completely skew in its indices. Contrast this to the field theory situation
where such a tensor does exist, namely the structure constants  $d_{\alpha \beta 
\gamma} $ of the `flat group' that is associated with the Scherk-Schwarz 
mechanism \cite{S-S}, and where indeed trilinear terms in $A_{\alpha} $ are present. 
From the viewpoint of finiteness, it is easy to see that such cubics are 
always proportional to the trace of the fermion mass matrix. But  the 
mass matrix calculated in (\ref{eq:27}), is always traceless ( which
follows from the fact that $\cA^{\alpha} $ and $\cB^{\beta} $ are off diagonal matrices \cite{GSO} ), 
and so provides further explanation of why such cubics are absent.
By contrast,  the trace of the fermion mass matrix is not generally zero in the 
field theory case \cite{SOFT1}.  
 
In conclusion, we have shown how specific dimension $\leq 3 $ operators appear in the 
low energy effective theory obtained from generalized Narain-type compactifications
of the heterotic string in which  the $N=4 $ local supersymmetry is spontaneously 
broken. In the global limit $\kappa \rightarrow 0 $,  one obtains explictly broken
$N=4 $ SYM theory, where the combination of supersymmetry breaking operators were
shown to be precisely those which preserve the ultraviolet properties of unbroken 
$N=4$ SYM. We have presented the simplest scenario, namely toroidal compactification
and no gauge symmetry breaking; a more general study will be presented elsewhere \cite{ISST}.
 The realization of the `stringy' GDR  mechanism considered here is 
particularly  suitable to generalizations that include orbifolds of
 Narain compactifications as stressed in \cite{GDRN}. It would be very interesting to compute the 
structure of the soft supersymmetry breaking terms in these theories as in the global limit,
they will provide examples of explicitly broken $N=2$ and $N=1$ theories.
(Note that this generalization has no obvious analogue within the usual 
GDR  of $d=10$ supergravities, which always yields spontaneously broken 
$N=4, d=4$ theories.) In this context, there has been interesting recent
work on the partial breaking of $N=4$ supersymmetry in string theory
\cite{partial}, giving rise to models with $N=2$ and $N=1$ supersymmetry.   
 A classification of
finiteness preserving soft terms within certain classes of (finite) $N=2 $ supersymmetric field theories
has been known for some time \cite{SOFT3}. Whether or not the results we have obtained here extend 
to these theories is worth pursuing. 
  
Finally, GDR  mechanisms have  been studied recently
in the context of M-theory  compactified on  $M_{10}\times  S^1 / Z_2 $ \cite{M1}.
These applications are  particularly interesting because the theory has bulk and boundary 
fields living in 11 and 10 dimensions. Since M-theory compactified on a line element is
 believed to describe strongly coupled $E_8 \times E_8 $ heterotic strings \cite{HW}, one 
might be able to investigate the effect of large string coupling on our results 
 by computing the soft terms obtained within the framework described in \cite{M1}.

\vskip1cm
{\bf\Large Acknowledgements} 
\vskip1cm
S.T. would like to thank M. Spalinski for useful discussions,
and the Royal Society of Great Britain for financial support.
I.S. was supported by PPARC.  

\end{document}